\documentclass{PoS}
\usepackage{upgreek}
\usepackage{bbm}
\usepackage[numbers,sort&compress]{natbib}
\usepackage{lipsum}

\usepackage{hyperref}% add hypertext capabilities
\newcommand{\slashed}[1]{\displaystyle{\not}{#1}}

\title{Heavy neutrinos in particle physics and cosmology}

\ShortTitle{Heavy neutrinos in particle physics and cosmology}

\author{Marco Drewes\\
  Technische Universit\"at M\"unchen, James Franck Stra\ss e 1, D-85748 Garching, Germany\\
        E-mail: {\rm marco.drewes@tum.de}}

\abstract{
Neutrinos are the only particles in the Standard Model of particle physics that have only been observed with left handed chirality to date. If right handed neutrinos exist, they would not only explain the observed neutrino oscillations, but could also be responsible for several phenomena in cosmology, including the baryon asymmetry of the universe, dark matter and dark radiation. A crucial parameter in this context is their Majorana mass, which in principle could lie anywhere between the eV scale and GUT scale. The implications for experiments and cosmology strongly depend on the choice of the mass scale. We review recent progress in the phenomenology of right handed neutrinos with different masses, focusing on scenarios in which the mass is at least a keV. We emphasise the possibility to discover heavy neutrinos that are responsible for the baryon asymmetry of the universe via low scale leptogenesis in near future experiments, such as LHC, BELLE II, SHiP, FCC-ee or CEPC.
}

\FullConference{The European Physical Society Conference on High Energy Physics\\
		22--29 July 2015\\
		Vienna, Austria}

\begin{document}
\noindent This is a very brief summary of the phenomenological implications of adding heavy right handed neutrinos to the Standard Model, based on my talk at HEP-EPS 2015. More details and a more complete list of references can e.g.\ be found in \cite{Drewes:2013gca,Drewes:2015jna}. I would like to apologise to all authors whose important contributions I had to omit here due to length restrictions.

\paragraph{Neutrino masses and the Seesaw Mechanism} - \label{part:seesaw}
The discovery of neutrino oscillations, which has been awarded the Physics Nobel Prize in 2015, remains the only confirmed proof of physics beyond the Standard Model (SM) that has been found in the laboratory. Explaining the oscillations by neutrino masses definitely requires the existence of new physical states. If neutrinos are Dirac particles, right handed (RH) neutrinos $\nu_L$ are required to construct a Dirac mass term $\bar{\nu}_Lm_\nu\nu_R$. 
If they are Majorana particles with a mass term $\overline{\nu_L} m_\nu \nu_L^c$, the only gauge invariant way to introduce such a term is via spontaneous symmetry breaking and higher dimensional operators, such as $\frac{1}{2}\bar{\ell_{L}}\tilde{\Phi}f\tilde{\Phi}^{T}\ell_{L}^{c}$ \cite{Weinberg:1979sa}.
Here $\ell_{L}=(\nu_{L},e_{L})^{T}$ are the LH lepton doublets and $\Phi$ is the Higgs doublet with $\tilde{\Phi}=\epsilon\Phi^*$, were $\epsilon$ is the antisymmetric $SU(2)$ tensor. 
The higher dimensional operators are not renormalisable and indicate that some particles with masses much above the typical energies of neutrino oscillation experiments have been ``integrated out''.
The probably simplest way to generate a neutrino mass term is adding $n$ RH neutrinos $\nu_R$ to the SM \cite{Minkowski:1977sc,GellMann:seesaw,Mohapatra:1979ia,Yanagida:1980xy,Schechter:1980gr,Schechter:1981cv}, which also seems well-motivated because all other fermions are known to exist with both chiralities.
In the following we focus on the scenario $n=3$. Since the $\nu_{R I}$ are gauge singlets, there exists no fundamental reason for this choice, except for the fact that the known fermions come in three generations. 
The most general renormalisable Lagrangian with only SM fields and $\nu_R$ reads  
\begin{eqnarray}
	\label{L}
	\mathcal{L} &=&\mathcal{L}_{SM}+ 
	i \overline{\nu_{R}}\slashed{\partial}\nu_{R}-
	\overline{\ell_{L}}F\nu_{R}\tilde{\Phi} -
	\tilde{\Phi}^{\dagger}\overline{\nu_{R}}F^{\dagger}l_L 
	-{\rm \frac{1}{2}}(\overline{\nu_R^c}M_{M}\nu_{R} 
	+\overline{\nu_{R}}M_{M}^{\dagger}\nu^c_{R}). 
	\end{eqnarray}
Here $\mathcal{L}_{SM}$ is the SM Lagrangian, 
The Majorana mass term $M_{M}={\rm diag}(M_1,M_2,M_3)$ introduces a new mass scale in nature,\footnote{Here $\nu_R^c=C\overline{\nu_R}^T$, the charge conjugation matrix is $C=i\gamma_2\gamma_0$ in the Weyl basis.} and $F$ is a matrix of Yukawa couplings. 
For $M_I> 1$ eV there are two distinct sets of mass eigenstates, 
which can be represent by flavour vectors of Majorana spinors $\upnu$ and $N$,\footnote{Here c.c. stands for the $c$-conjugation defined above.} %. The elements $\upnu_i$ of the vector 
\begin{equation}
\upnu=V_\nu^{\dagger}\nu_L-U_\nu^{\dagger}\theta\nu_{R}^c +{\rm c.c.} \ , \ N=V_N^\dagger\nu_R+\Theta^{T}\nu_{L}^{c} +{\rm c.c.}
\end{equation}
$V_\nu$ is the usual neutrino mixing matrix $V_\nu\equiv (\mathbbm{1}-\frac{1}{2}\theta\theta^{\dagger})U_\nu$
with  
$\theta\equiv m_D M_M^{-1}$, $m_D\equiv Fv$. 
$U_\nu$ is its unitary part, $V_N$ and $U_N$ are their equivalents in the sterile sector and $\Theta\equiv\theta U_N^*$. The elements $\upnu_i$ of the vector $\upnu$ are mostly superpositions of the ``active'' SU(2) doublet states $\nu_L$ and have light masses $\sim -F^2 \times v^2/M_I\ll M_I$. 
The elements $N_I$ of $N$ are mostly superpositions of the ``sterile'' singlet states $\nu_R$ with masses of the order of $M_I$. 
At energies below the electroweak scale, the \emph{heavy neutral leptons} $N_I$ interact with the SM via mixing with active neutrinos, which is quantified by the matrix elements $\Theta_{\alpha I}$. 
The behave just as SM neutrinos, but with a larger mass $M_I$ and a cross sections suppressed by factors $U_{\alpha I}^2\equiv|\Theta_{\alpha I}|^2\ll 1$.
The unitary matrices $U_\nu$ and $U_N$ diagonalise the mass matrices 
\begin{eqnarray}
m_\nu\simeq-v^2FM_M^{-1}F^T=-\theta M_M \theta^T \label{activemass} \ , \
M_N\simeq M_M + \frac{1}{2}\big(\theta^{\dagger} \theta M_M + M_M^T \theta^T \theta^{*}\big), 
\end{eqnarray}
respectively. If (\ref{activemass}) is the only source of neutrino masses, $n$ must at least equal the number of non-zero $m_i$, i.e.\ $n\geq 2$ if the lightest neutrino is massless and $n\geq 3$ if it is massive. Phenomenologically, the most interesting properties of the $N_I$ are their masses $M_I$ and interaction strengths
\begin{eqnarray}
U_{\alpha I}^2\equiv |\Theta_{\alpha I}|^2 \ , \ U_I^2\equiv\sum_\alpha U_{\alpha I}^2 \ , \ U_\alpha^2\equiv\sum_I U_{\alpha I}^2.
\end{eqnarray}

The new particles $N_I$ may be responsible for several phenomena in cosmology, including the observed Dark Matter (DM) and the baryon asymmetry of the universe (BAU).
The role they play in particle physics and cosmology strongly depends on the choice of the new mass scale(s) $M_I$, see \cite{Drewes:2013gca,Drewes:2015jna} for a review.
Neutrino oscillation data  and (\ref{activemass}) only constrain the combination $FM_M^{-1}F^T$ of mass and coupling, leaving much freedom to vary $M_I$.
If the $F$ are perturbatively small, then $M_I$ should be at least 1-2 orders of magnitude below the Planck scale. On the lower end the $M_I$ can be as small as a few eV \cite{deGouvea:2005er,deGouvea:2009fp}.\footnote{If all $M_I=0$, then $\upnu_i$ are Dirac neutrinos, which remains to be a possibility that is consistent with all known data. Small values $0<M_I\ll 1$ eV are, however, mostly excluded \cite{deGouvea:2009fp}.} For $n\geq3$ any value in between is experimentally allowed, though there is a clear preference for $M_I>100$ MeV if the $N_I$ are required to be the only source of active neutrino masses \cite{Hernandez:2014fha}.

\paragraph{The GUT-seesaw} - 
In the probably most studied version of the seesaw mechanism the $M_I$ are far above the electroweak scale. 
This choice is primarily motivated by aesthetic arguments: For values $F_{\alpha I}\sim 1$, neutrino masses near the Planck limit $\sum_i m_i< 0.23$ \cite{Ade:2013zuv} imply values $M_I\sim 10^{14}-10^{15}$ GeV, slightly below the suspected scale of grand unification. 
This scenario can easily be embedded in grand unifying theories. 
For $M_I\gtrsim4\times10^{8}$ GeV \cite{Davidson:2002qv}, the CP-violating decay of $N_I$ particles in the early universe can furthermore explain the BAU via leptogenesis \cite{Fukugita:1986hr}.
Flavour effects \cite{Barbieri:1999ma,Abada:2006ea,Abada:2006fw,Nardi:2006fx,Blanchet:2006be} can reduce this lower bound by $1-2$ orders of magnitude \cite{Antusch:2009gn}.\footnote{The consistent description of all quantum and flavour effects remains an active field of research \cite{Buchmuller:2000nd,Anisimov:2008dz,Anisimov:2010aq,Beneke:2010dz,Anisimov:2010dk,Beneke:2010wd,Drewes:2010pf,Frossard:2012pc,Drewes:2012qw,Garny:2011hg,Drewes:2012ma,Dev:2014laa,Dev:2014wsa}.}
If the $M_I$ are indeed that large, then $N_I$ cannot be found in any near future experiment (and possibly never). The only trace they leave in experiments can be parametrised in terms of higher dimensional operators in an effective Lagrangian \cite{Broncano:2002rw}, which can e.g.\ be constrained by searches for rare processes.
On the positive side this means that one can get some information about physics at very high scales. For a degenerate $M_I$-spectrum or $n=1$, the resulting bounds on $N_I$-properties can indeed be quite strong, see e.g. \cite{Blennow:2010th}, for $n=3$ they are much weaker \cite{Drewes:2015iva}. 
On the negative side, the seesaw mechanism is not the only way to generate these operators, and without directly finding the new states, it is impossible to know their origin.

\begin{figure}
\begin{center}
\includegraphics[width=0.7\textwidth]{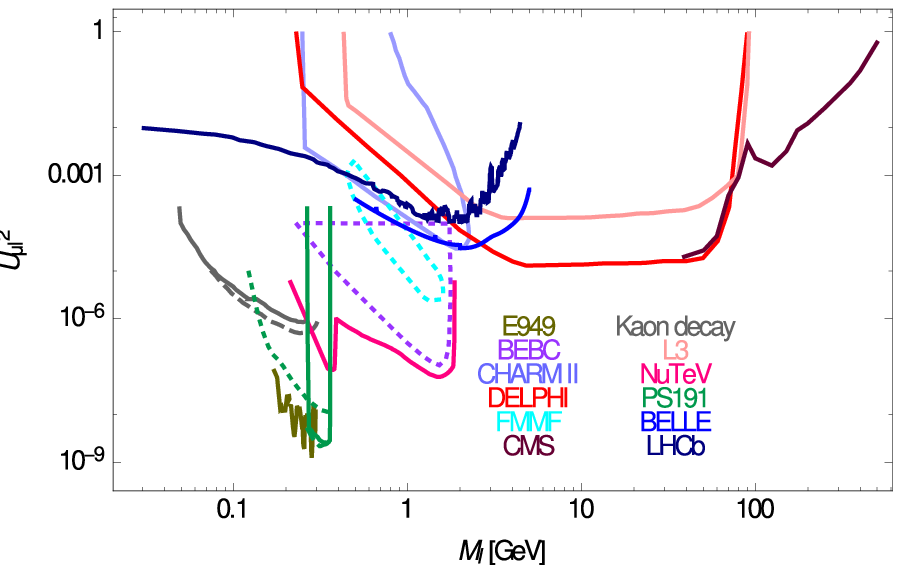}\\
\includegraphics[width=0.7\textwidth]{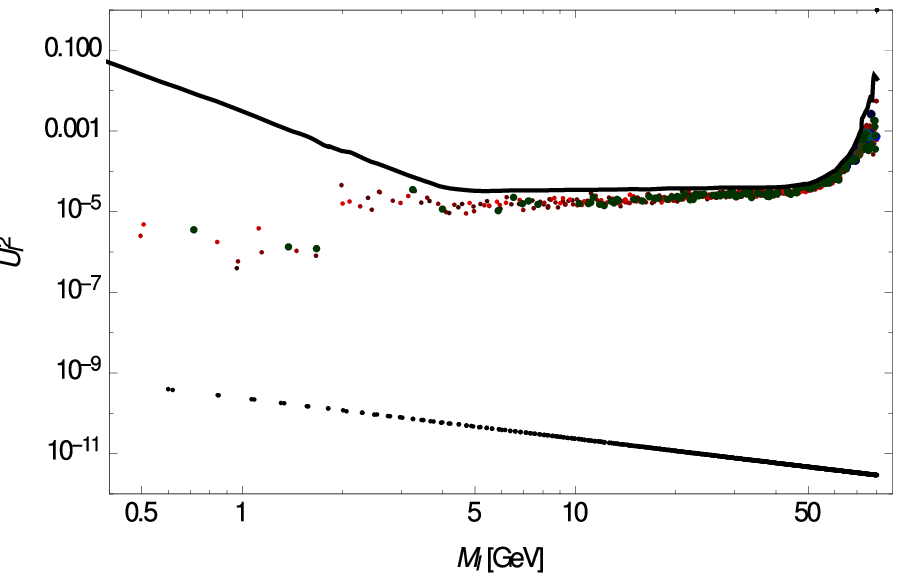}
\caption{\label{UmuPlot}
\emph{Upper panel}: Constraints on $U_{\mu I}^2$ from the experiments DELPHI \cite{Abreu:1996pa}, L3 \cite{Adriani:1992pq}, LHCb \cite{Aaij:2014aba}, BELLE \cite{Liventsev:2013zz}, 
BEBC \cite{CooperSarkar:1985nh}, %as in Pascoli
FMMF \cite{Gallas:1994xp}, %as in Pascoli
E949 \cite{Artamonov:2014urb},
PIENU \cite{PIENU:2011aa}, 
TRIUMF/TINA \cite{Britton:1992xv},
PS191 \cite{Bernardi:1987ek},  
CHARMII \cite{Vilain:1994vg},
NuTeV \cite{Vaitaitis:1999wq}, 
NA3 \cite{Badier:1985wg},
CMS \cite{Khachatryan:2015gha}
and kaon decays in \cite{Yamazaki:1984sj,Hayano:1982wu}.
\emph{Lower panel}: The black dots indicate the upper and lower bound on the sum $U_I^2=\sum_\alpha U_{\alpha I}^2$ from collider searches and neutrino oscillation data with $\sum_im_i=0.23$ eV and inverted hierarchy (chosen as an example).
The colourful dots show the largest possible value of $U_I^2$ for given $M_I$ found in a Monte Carlo scan with $10^8$ points that is consistent with these bounds as well as indirect searches (rate lepton decays, electroweak precision data, lepton universality, neutrinoless double $\beta$-decay searches and CKM unitarity constraints).
 The colour indicates the most saturated bound (green is lepton universality, red neutrinoless double $\beta$-decay), the shade indicates the degree of saturation of that bound. Details will be given in an updated version of \cite{Drewes:2015iva}.
}
\end{center}
\end{figure}

\paragraph{The TeV- and electroweak seesaw} - 
The highest scale that can be probed directly by experiments is the TeV scale. 
Searches for $N_I$ have been undertaken at the ATLAS and CMS experiments at the Large Hadron Collider (LHC) \cite{ATLAS:2012ak,Khachatryan:2014dka,Khachatryan:2015gha}, so far without positive result, see Fig.~\ref{UmuPlot}. 
The experimental challenge lies the smallness of the Yukawa interactions $F$; a relatively low seesaw scale $M_I$ in (\ref{activemass}) generically requires small values of the $F_{\alpha I}\sim F_0\equiv(m_i M_I/v^2)^{1/2}$, hence tiny branching ratios. 
In the minimal seesaw (\ref{L}) a discovery at the LHC is only realistic if some individual $F_{\alpha I}$ are much bigger than $F_0$, and the smallness of the $m_i$ is achieved due to a cancellation in (\ref{activemass}) \cite{Kersten:2007vk,Ibarra:2011xn,Das:2012ze}, e.g.\ due to an approximate conservation of lepton number \cite{Schechter:1980gr,Mohapatra:1986aw,Mohapatra:1986bd,Gavela:2009cd}. Chances are generally better in extensions of (\ref{L}) in which the $N_I$ have additional interactions, see \cite{Mohapatra:1986bd,Chikashige:1980ui}.
Direct searches at LHC and future colliders have e.g.\ been studied in  \cite{Blondel:2014bra,Aaij:2014aba,Das:2014jxa,Asaka:2015oia,Antusch:2015mia,Antusch:2015rma,Ng:2015hba,Izaguirre:2015pga,Graverini:2015dka,Hung:2015hra,Peng:2015haa,Das:2015toa,Dev:2015kca}.
Meanwhile, the parameter space can already be constrained by indirect means, including rare decays \cite{Ibarra:2011xn,Abada:2014kba,Drewes:2015iva}, lepton universality \cite{Abada:2013aba,Basso:2013jka,Endo:2014hza,Asaka:2014kia,Drewes:2015iva}, neutrinoless double $\beta$-decay \cite{Atre:2009rg,Bezrukov:2005mx,Ibarra:2011xn,Lopez-Pavon:2015cga,Drewes:2015iva}, electroweak precision data \cite{delAguila:2008pw,Atre:2009rg,Akhmedov:2013hec,Blas:2013ana,Drewes:2015iva}, CKM unitarity \cite{Atre:2009rg,Antusch:2014woa,Drewes:2015iva} or lepton flavour violation in muonic systems \cite{Abada:2015oba}.

$N_I$ with $M_I$ at the electroweak or TeV scale are also interesting cosmologically because they can generate the BAU via resonant leptogenesis either during their decay \cite{Pilaftsis:2003gt}  or thermal production ("baryogenesis from neutrino oscillations") \cite{Akhmedov:1998qx,Asaka:2005pn,Garbrecht:2014bfa}. 

\paragraph{The GeV-seesaw} - 
If the $M_I$ are  below the masses of the W and Z-bosons, then $N_I$ can be produced in the decay of gauge bosons. This allows to impose much stronger constraints \cite{Abreu:1996pa,Graverini:2015dka,Antusch:2015mia,Drewes:2015iva,Abada:2015zea,Dib:2015oka,Helo:2013esa,Izaguirre:2015pga,Gago:2015vma}.
For $M_I$ below the mass of the B-mesons (and even more below the D-meson mass), the existing direct and indirect experimental constraints are considerably stronger \cite{Atre:2009rg,Asaka:2015oia,Drewes:2015iva}, see Fig.~\ref{UmuPlot}. 
On one hand the $N_I$ can be produced efficiently in meson decays \cite{Gorbunov:2007ak,Atre:2009rg,Boyarsky:2009ix,Ruchayskiy:2011aa,Canetti:2014dka,Drewes:2015iva}.
On the other hand neutrino oscillation data impose stronger bounds on the sum $U^2_I\equiv\sum_\alpha U_{\alpha I}^2$ \cite{Canetti:2012kh,Hernandez:2014fha,Drewes:2015iva}. 
Finally, the requirement to decay before the formation of light elements in big bang nucleosynthesis can impose a lower bound  on $U^2_I$ as a function of $M_I$ \cite{Dolgov:2003sg,Ruchayskiy:2012si,Hernandez:2014fha}.
Combining all these bounds allows to identify the phenomenologically allowed range of $N_I$ parameters \cite{Drewes:2015iva}, see Fig.~\ref{UmuPlot}.

The BAU can be explained for $M_I\sim$ GeV via leptogenesis during the thermal production of the $N_I$  \cite{Akhmedov:1998qx,Asaka:2005pn,Canetti:2010aw,Shuve:2014zua,Drewes:2012ma,Canetti:2012vf,Canetti:2014dka,Abada:2015rta,Hernandez:2015wna}. This requires a degeneracy in the masses $M_I$ for $n=2$ \cite{Canetti:2012vf,Shuve:2014zua}, but no degeneracy is needed for $n=3$ \cite{Drewes:2012ma}. In a small fraction of the parameter space the CP-violation responsible for the BAU comes from the phases in $U_\nu$ that may be measured in neutrino oscillation experiments \cite{Canetti:2012vf}, but in general it lies in the sterile sector and can only be measured in $N_I$ decays if their mass spectrum is degenerate \cite{Cvetic:2014nla}.

The leptogenesis parameter space will be further explored in the near future. 
For $M_I$ below the D-meson mass this is e.g.\ done by the NA62 experiment \cite{Moulson:2013oga}, for heavier masses LHCb and BELLE II will improve the bounds \cite{Canetti:2014dka}. 
Also searches at T2K \cite{Asaka:2012bb} or with DUNE (formerly LBNE) \cite{Adams:2013qkq} have been proposed. The most significant improvement could be made with the proposed SHiP experiment \cite{Bonivento:2013jag,Alekhin:2015byh} or a future lepton collider \cite{Blondel:2014bra,Antusch:2015mia}.

\paragraph{The keV-seesaw}
Sterile neutrinos $N_I$ are massive, feebly interacting and can be very long lived.
This makes them obvious DM candidates \cite{Dodelson:1993je,Shi:1998km}.

Observations with the Astro-H satellite \cite{Takahashi:2012jn} may help to clarify the situation. Since thermal production via mixing is unavoidable \cite{Dodelson:1993je}, an upper bound on $U_I^2$ can also be obtained from the requirement not to produce too much DM. 
The DM mass $M_I$ is bound by phase space considerations to be larger than a keV \cite{Gorbunov:2008ka}. $U_I^2$ can also be constrained in the laboratory by KATRIN-type experiments \cite{Mertens:2014nha}.
Finally, the free streaming length of DM in the early universe can be constrained by the effect it has no structure formation.
The way how this can be translated into a bound on the sterile neutrino mass depends on the way they were produced in the early universe \cite{Abazajian:2001nj,Boyarsky:2008xj,Lovell:2011rd,Kusenko:2009up,Boyarsky:2009ix}. 
A minimal population is produced thermally via mixing \cite{Dodelson:1993je}, which, however, depends on the chemical potentials in the primordial plasma via the MSW effect \cite{Shi:1998km,Ghiglieri:2015jua,Venumadhav:2015pla}. If this population composes all the DM, then structure formation implies  $M_I>3.3$ keV \cite{Viel:2013apy}.
Sterile neutrino DM may also be produced in the decay of a scalar field \cite{Shaposhnikov:2006xi,Kusenko:2006rh,Petraki:2007gq,Bezrukov:2009yw,Adulpravitchai:2014xna,Merle:2013wta,Frigerio:2014ifa,Merle:2015oja,Haba:2014taa,Molinaro:2014lfa,Adulpravitchai:2015mna,Lello:2014yha,Drewes:2015eoa}, or from gauge interactions, in which case a dilution at later time is necessary to avoid a too large DM density \cite{Nemevsek:2012cd,Bezrukov:2009th,Bezrukov:2012as,Patwardhan:2015kga}.
In these scenarios the initial momentum distribution of the DM is different, leading to different bounds \cite{Merle:2014xpa}.

\paragraph{Conclusion} - 
To date, neutrino oscillations remain to be the only established piece of evidence for the existence of physics beyond the SM that has been found in the laboratory. They can easily be explained if the SM is complemented by heavy RH neutrinos. These new particles could also be responsible for unexplained phenomena in cosmology, including the DM and baryon asymmetry of the universe. If their masses are below the TeV scale, they may be found in near future experiments. \\
 
\paragraph{Acknowledgements} - I would like to thank my collaborators Laurent Canetti, Bj\"orn Garbrecht and Mikhail Shaposhnikov, who contributed to some of the results summarised here. This work was supported by the Gottfried Wilhelm Leibniz program of the Deutsche Forschungsgemeinschaft (DFG) and the DFG cluster of excellence Origin and Structure of the Universe.\\

\paragraph{References}

%\bibliographystyle{JHEP}
%\newpage
%\section{References}
%REFERENCES TOGETHER AS IN NUFACT!
\begin{footnotesize} 
\bibliographystyle{apsrev}
%\bibliography{all}
%\printbibliography
\vspace{-0.5cm}
\begingroup
\renewcommand{\section}[2]{}
   \setlength{\bibsep}{1pt}
    %\setstretch{1}
    %{\fontsize{10}{12} 
      \bibliography{all}
    %}
\endgroup

%\vspace{0.2cm}
%\noindent%\footnotesize{
%\noindent Numbers xxxx.xxxx refer to \emph{arxiv.org} identifiers.
%}
\end{footnotesize}

%\section{...}
%\begin{thebibliography}{99}
%\bibitem{...}
%....
%\end{thebibliography}

\end{document}